\documentclass[aip,apl,reprint,groupedaddress]{revtex4-1}
\usepackage{graphicx}
\begin{document}


\title{
Effect of thermal inhomogeneity for THz radiation from intrinsic Josephson junction stacks
of Bi$_2$Sr$_2$CaCu$_2$O$_{8+\delta}$
}



\author{Itsuhiro Kakeya}
\email[]{kakeya@kuee.kyoto-u.ac.jp}
\affiliation{Department of Electronic Science and Engineering, Kyoto University,
Nishikyo-ku, Kyoto 615-8510, Japan}

\author{Yuta Omukai}
\affiliation{Department of Electronic Science and Engineering, Kyoto University,
Nishikyo-ku, Kyoto 615-8510, Japan}

\author{Minoru Suzuki}
\affiliation{Department of Electronic Science and Engineering, Kyoto University,
Nishikyo-ku, Kyoto 615-8510, Japan}

\author{Takashi Yamamoto}
\thanks{Present address: Quantum Beam Science Directorate, Japan Atomic
Energy Agency, Takasaki, Gunma 370-1292, Japan.}
\affiliation{Institute of Materials Science, University of Tsukuba,
Tsukuba, Ibaraki 305-8573 Japan}
\author{Kazuo Kadowaki}
\affiliation{Institute of Materials Science, University of Tsukuba,
Tsukuba, Ibaraki 305-8573 Japan}


\date{\today}

\begin{abstract}
Terahertz radiation from the mesa structures of Bi$_2$Sr$_2$CaCu$_2$O$_{8+\delta}$ is detected
in samples with thin electrodes $< 100$ nm.
In samples with thick electrodes $\simeq$ 400 nm, neither radiations nor voltage jumps in current-voltage characteristics are detected.
This suggests that the thin electrode helps excite the Josephson plasma oscillation as a result of the poor heat flow
through the electrode.
The shielding effect by the electrode is not essential.
We consider that the local temperature rise is the origin of the synchronization of the phase kink for terahertz radiation.
\end{abstract}


\maketitle


A single crystal of Bi$_2$Sr$_2$CaCu$_2$O$_{8+\delta}$ (Bi2212) is described by a stack of atomic-scale Josephson junctions
referred to as intrinsic Josephson junctions (IJJs) with CuO$_2$ double layers being superconducting electrodes and SrO-Bi$_2$O$_2$-SrO layers being barrier layers~\cite{Kleiner1992}.
It has been expected that the synchronization of the Josephson plasma oscillations excited inside
the IJJs leads to intensive emission of terahertz (THz) electromagnetic waves because the Josephson plasma oscillation is excited with very little damping inside the large superconducting energy gap ($\Delta \sim$ 40 meV)~\cite{Tachiki:1994,Suzuki2000}.
The methods for the synchronization for coherent THz emission have been discussed in terms of the dynamics of the Josephson vortex lattice~\cite{Machida2000,Tachiki:2005}.
Although some groups have detected THz waves from IJJs~\cite{Batov2006,Bae:2007}, the highly specialized techniques used there do not allow followers to reproduce the results.

Recently, it was observed that monochromatic and continuous electromagnetic waves with frequencies between 0.3 and 0.9 THz and powers up to 5 $\mu$W
were radiated from rectangular IJJ mesas formed on surfaces of single crystals of Bi2212~\cite{Ozyuzer2007,Minami2009}.
The radiation is explained as the half-wavelength cavity resonances of IJJs included in the mesa.
The standing waves of the Josephson plasma oscillations are synchronized along the $c$-axis, although the length of each IJJ is varied due to the trapezoidal cross section of the mesa.
The excitation frequency is determined by the matching between the cavity length and the AC Josephson effect in the IJJs.
The numerical simulations by several authors reproduced that the formation of the dynamical phase variation along the mesa edge ($\pi$-kink) yields far-field electromagnetic THz radiation~\cite{Lin2008,Koshelev2008a,Koyama2009}.
Subsequent experimental results on the radiation from square and cylinder-shaped mesas~\cite{Tsujimoto2010,Kashiwagi2011} are explained based on this idea
with degeneracies of the cavity modes and the partial contribution of the stacked IJJs for the radiation.
Wang {\it et al.} argued that the temperature dependence of the emission frequency is attributed to the change in the effective cavity geometry
through the observation of the temperature inhomogeneity in the mesa, in which non-superconducting (hotspot) regions are found under the bias for the emission~\cite{Wang2010}.
The relation between the temperature distribution and the formation of the $\pi$-kink aligned along the $c$-axis is the central concern of this paper.


Here, we present the data of the bolometric detection for the THz electromagnetic waves from Bi2212 mesa structures with three different electrode thicknesses.
Strong emissions of electromagnetic waves were detected in the mesas with thinner silver electrodes,
but no bolometer response was observed for the mesas with the thickest electrodes.
This implies that the temperature rise due to self-heating induces the synchronization of the transverse Josephson plasma oscillations~\cite{Kakeya1998} excited in each IJJ.



\begin{figure}[tb]
\includegraphics[width = \linewidth]{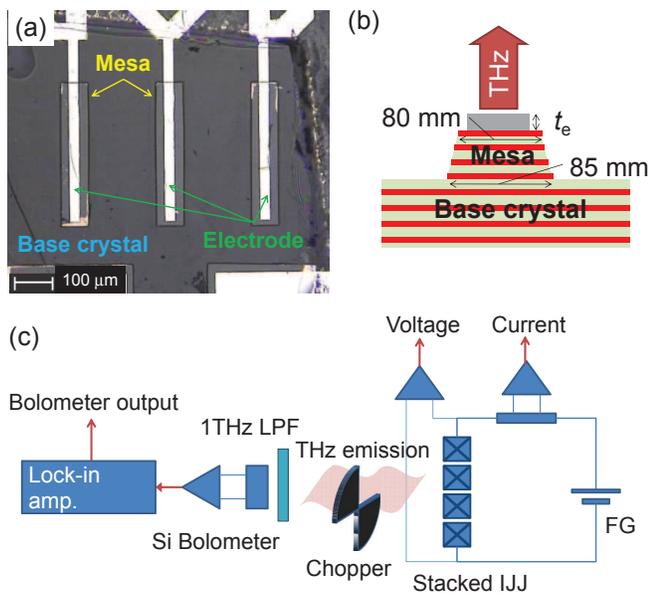}%
\caption{
(a) Microscope image of a device, shoing mesas and upper electrodes.
(b) Schematic illustration of the cross section of the mesa structure.
(c) Diagram of detection system.
}
\label{experimental}
\end{figure}

The devices used in this study were prepared by the following procedure.
A Bi2212 single crystal
grown by the TSFZ method was annealed at 650 $^{\circ}$C for 75 hours in an Ar atmosphere to reduce its
 oxygen concentration.
The crystal was cut into small pieces approximately 1 $\times$ 1 $\times$ 0.05 mm$^3$, one of which was glued onto a sapphire substrate
with Stycast 1266 epoxy resin.
To decrease the contact resistance between electrodes and a crystal,
the single crystal was cleaved in a vacuum and a 30-nm thick Ag film (lower electrode) was immediately evaporated
on its surface.
A couple of mesa structures were fabricated on a Bi2212 single crystal by photolithography
and Ar ion milling, and the geometries of the mesas are the same as $80 \times 400 \times 1.2$ $\mu$ m$^3$ (Fig. \ref{experimental}(a)).
Since a single IJJ is 1.5 nm thick, the mesa comprises approximately 800 IJJs.
The mesas are trapezoidal shaped in which the lengths of the top sides are approximately 5 $\mu$m shorter than those of the bottom sides (Fig. \ref{experimental}(b)).
Then, additional upper Ag layers were evaporated and patterned into the upper electrode, which extends to the outside of the mesa.
The thickness of the upper electrode $t_e$ is an important parameter in this study.
As listed in in Table \ref{Table}, $t_e$ was varied as 30, 70, and 400 nm for device types A, B, and C, respectively.
The width of all the electrodes was 30 $\mu$m.
Note that mesa B-1 was fabricated from C-1; mesa B-1 was formed on the cleaved surface of the crystal formerly used for a C-type device.

The devices were mounted on the sample stage of an Oxford CF1204 helium flow cryostat. A mesa was biased by triangular wave voltage of $\sim$ 1 mHz.
The THz radiation from the mesa was chopped with a frequency of $\sim$40 Hz and detected with
a Si-bolometer (Infrared Lab.) with a 1-THz low-pass filter.
The detector window was parallel to the $c$-axis in most cases.
The output signal of the bolometer was amplified with
a high-pass 46 dB preamplifier and a lock-in amplifier before acquired by a computer (Fig. \ref{experimental}(c)).
The detection limit was approximately 0.01 nW.

\begin{table}[tb]
\caption{
Sample used in this study and their values for $t_e$, $T_c$, $I_c$ at $\sim$ 20 K, the voltage and temperature ranges where THz radiation was observed.
A-2a and 2b are on the same crystal chip. No emission was observed in the two C-type devices.
}
\label{Table}
\begin{tabular}{l c c c c c c }
\hline
 & $t_e$ [nm] & $T_c$ [K] & $I_c$ [mA] & Voltage [V] & Temp. [K] \\
\hline
A-1 & 30 & 89 & 42 & 0.95--0.74 &20--45 \\
A-2a &  30 & 83 & 42 & 0.76--0.28 & 20--44 \\
A-2b &  30 & 83 & 43 & 0.75--0.29 & 20--44 \\
B-1 &  70 & 91 & 30 & 0.90--0.64 & 20--66 \\
C-1,-2 & 400 & 89 & 60 & N/A & N/A \\
\hline
\end{tabular}
\end{table}

\begin{figure}[tb]
\includegraphics[width=\linewidth]{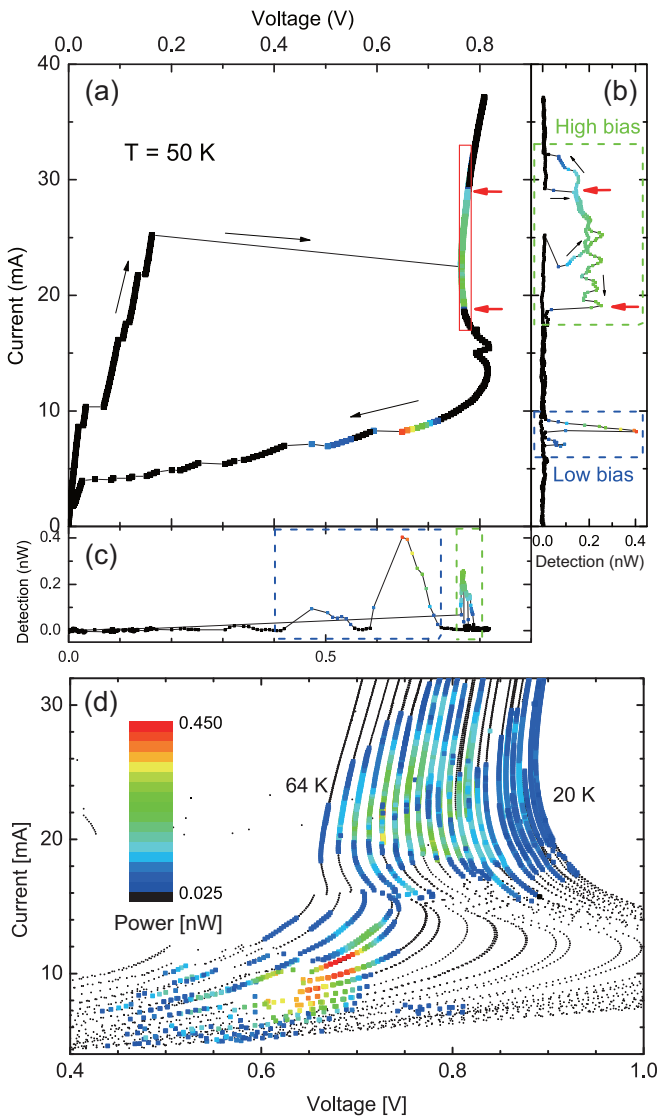}
\caption{
$I-V$ characteristics (a) and bolometer response in B-1 at 50 K as a function of either bias current (b) or voltage (b).
In (d), $I-V$ characteristics of B-1 at temperatures from 20 to 64 K with an interval of 2 K are plotted.
For all panels, colored symbols represent bolometric detections of THz wave with the detected power being scaled by color-bar in the panel (d).
A red rectangle in (a) corresponds to the axes of the blow-up plot shown in Fig. \ref{data_B+C}(a).
Thick red arrows indicate onset and end of the high-bias emission when bias is decreased.
}
\label{rawdata}
\end{figure}
\begin{figure}[tb]
\includegraphics[width=\linewidth]{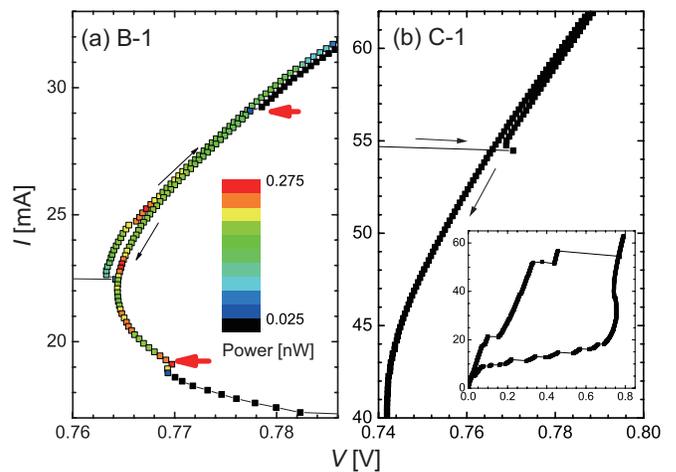}
\caption{
Blow-up plots of quasiparticle branch of $I-V$ characteristics in mesas B-1 (a) and C-1 (b) at 50 K.
Small voltage steps shown by red arrows are found at onset and offset of emission whereas no voltage step was found in C-1.
Inset in (b) is entire $I-V$ characteristics of C-1.
Colors in (a) represent detected intensity as scaled by color-bar.
}
\label{data_B+C}
\end{figure}

The THz emission was observed in the mesas of $t_e$ = 30 and 70 nm, whereas no bolometer response was observed in the mesas of $t_e=400$ nm.
Figure \ref{rawdata} represents typical current-voltage ($I-V$) characteristics and bolometer responses in mesa B-1,
in which THz emission was detected in two bias regions.
One is in the low bias region below 10 mA, where the $I-V$ curve frequently shows discontinuity due to the retrapping phenomena.
The emission in this region closely resembles the one originally reported by Ozyuzer {\it et al.}~\cite{Ozyuzer2007}
The other is in the high bias region above 15 mA, where the $I-V$ curve is almost vertical and shows bending forward.
The emission was observed continuously in a current range between 18 and 30 mA both with the increasing and decreasing current.
Note that small voltage steps in the $I-V$ curve (Fig. \ref{data_B+C}(a)) were found at the onset and the end of emission, as indicated by thick red arrows in Fig. \ref{rawdata}(b).
The emission intensity as a function of the bias current shows a broad maximum with reproducible small peaks as shown in Fig. \ref{rawdata}(b).
The difference between the two emissions is highlighted in Fig. \ref{rawdata}(d), where $I-V$ curves at various temperatures are plotted with colors indicating the detected THz intensity.
The high bias emission is observed in a broad optimum $I-V$ area centered at $(I, V)\simeq $ (25 mA, 0.75 V),
whereas the low bias emission is observed in a sharp maximum at $(I, V)\simeq $ (10 mA, 0.68 V) with remarkably asymmetric broadening with respect to voltage.
FT-IR spectroscopy of the high bias emission from B-1 at 35 K reveals that the emission frequency is 0.516 THz with the spectrum width being 0.75 GHz, the instrumental limit.
Details of the temperature dependence of their spectra are discussed elsewhere.

These experimental results provide an important finding that the emission of THz waves is detected in the mesas with $t_e <$  100 nm.
This agrees with all previous reports on THz emission, {\em i.e.}, $t_e \simeq 100$ nm~\cite{Ozyuzer2007,Minami2009} and  30 nm~\cite{Wang2010}.
If we notice that the emitting device is also formed after cleaving the non-emitting device on the same crystal,
then the most remarkable difference between the emitting and non-emitting devices is the electrode thickness.
Here, we consider two scenarios to explain why the THz emission is not observed from the mesas with $t_e$= 400 nm electrodes;
(a) the synchronized Josephson plasma oscillations are {\em not excited} inside the mesa due to the lack of thermal inhomogeneity and
(b) the synchronized Josephson plasma oscillations are {\em excited but shielded} by the thick electrode.
In the following paragraphs, we describe validity of the first scenario and invalidity of the second scenario.

To elucidate the validity of the first scenario, we first argue that temperature inhomogeneity is less remarkable in a mesa with a thicker electrode.
The bias current in the voltage state heats the mesa by the Joule heating.
The induced heat dissipates into the sapphire substrate and the sample stage underneath the mesa.
Since the $c$-axis thermal conductivity of Bi2212 was reported as 0.5 W/mK at 50 K~\cite{Crommie1991}, which is three orders smaller than that of Ag,
the main heat flow path is not via the base crystal but via the electrode attached to the mesa in this study ($t_e>$1/100 of the mesa thickness).
Considering that the electrode works as a primary heat-flowing channel, the electrode thickness will seriously influence the mesa temperature.
The temperature rise as a result of the self-heating is considered more pronounced and its temperature distribution is supposed to be more inhomogeneous in a thicker mesa with a thinner electrode.
It is also presumed that the temperature distribution is more uniform in a mesa with a thicker electrode because the Ag electrode over the mesa helps the temperature of the mesa uniform.
Thus the temperature distributions of the C type mesas are less inhomogeneous than those of the A and B type mesas.

Here, we suggest that the temperature inhomogeneity stimulates the THz emission via the formation of the  $\pi$ kinks aligned along the $c$-axis.
Numerical studies indicate that the synchronization of the Josephson plasma oscillations yielding far-field THz electromagnetic waves is given by the aligned $\pi$-kinks.
Nevertheless, it is energetically unfavorable to induce such a steep phase variation in an IJJ in the equilibrium state\cite{Tachiki2009}.
It has been argued that the $\pi$-kink is attributed to the inhomogeneous distribution of the critical current density\cite{Koshelev2008a} and the nucleations of Josephson vortex-antivortex pairs\cite{Krasnov2011a}.
Both are related to the temperature distribution inside the mesa because the vortices are attracted to the local minima of the superconducting order parameter.
Yurgens pointed out that the temperature distribution of the mesa under the radiating conditions is quite inhomogeneous through numerical calculations; the center of the mesa locally exceeds $T_c$ while the rest is still superconducting at the negative resistance state~\cite{Yurgens2011}.
Therefore we believe that the $\pi$-kink formation is attributed to the local temperature rise inside the mesa (another {\em hotspot}), which is remarkable in the mesas with $t_e<100$ nm in this study.
In mesas with $t_e$ = 400 nm, the local temperature rise is too small to excite the aligned $\pi$-kinks inside the mesa.

Other various experimental results can be explained based on this scenario.
The local temperature rise corresponds to the local suppression of $c$-axis Josephson critical current $J_c$.
Koyama {\it et al.} figured out that the asymmetric variation of $J_c$ along the $x$-direction (short edge of the mesa) induces the $\pi$-kink
and results in THz emission with much higher intensity than homogeneous $J_c$; 10\% inhomogeneity
yields an enhancement in emission by two orders of magnitude~\cite{Koyama2011}.
Such asymmetric distribution of $J_c$ along the short edge will be expected for an asymmetric temperature distribution
if we consider accidental asymmetric placement of the electrode on the mesa, which is inevitable for the practical device fabrication.
Indeed, another type of emission with strong intensity uniquely observed in A-1 is given by an unstable bias condition; emitting bias voltage and current vary from scan to scan.
A tiny asymmetry of temperature distribution triggers the $\pi$-kink aligned along the $c$-axis,
which gives synchronized Josephson plasma oscillation.
This interpretation is consistent with the poor reproducibility of the $I-V$ curve and the THz emission among
previous results.

The interpretation for the poor reproducibility enables us to explain the non-monotonic temperature dependence of the emission intensity, indicating the maximum around 30-40 K~\cite{Kadowaki2008}.
It is widely known that the temperature dependence of $J_c$ is more significant at higher temperature.
Consider hotspot $\Delta T$, which is higher than the whole mesa at temperature $T$, and the suppression of critical current $\Delta J_c$ that depends on $T$,
a constant $\Delta T$ would give larger $\Delta J_c$ at higher $T$.
At low temperatures where no THz emission is observed, $\Delta J_c$ is too small to induce the aligned $\pi$-phase kinks.
With increasing temperature, $\Delta J_c$ for the same $\Delta T$ becomes larger while the synchronization among the IJJs is smeared.
The balance between $\Delta J_c$ and the coupling between IJJs may give optimum emission power around 30-40~K.

Next, we argue that the second scenario is invalid.
The skin depth of a metal is described as
$
\delta=\sqrt{2 \rho / \omega \mu \sigma},
$
where $\rho$ and $\mu$ are its resistivity and permeability.
The physical parameters which may reproduce this situation, $\rho = 0.3$ $\mu\Omega$ cm (77 K), $\mu=4 \pi \times 10 ^{-7}$ H/m, and $\omega/2\pi=0.6$ THz,
would give the damping factor by the 400-nm thick electrode as $e^{-11} \sim 10^{-5}$ at low temperatures $\sim 77$ K if the shielding effect was taken into consideration.
However,
supposing that the $\pi$-kink state was realized, the contribution to the far field THz detection from the uncovered parts of the mesa would be {\em not} negligible because the uncovered parts occupy the area more than half of the surface of the mesa as seen in Fig. \ref{experimental}(a).
Moreover, no voltage jumps implying the THz excitation was found in the $I-V$ curve indicated in Fig. \ref{data_B+C}(b).
This is a sharp contrast to the case of the mesa B-1 shown in Fig. \ref{data_B+C}(a).
Very recently, similar THz radiations from an IJJ stack sandwiched by two superconducting electrodes (double-sided sample) were reported~\cite{Wang2011}.
This experimental fact also excludes the validity of this scenario because the superconducting electrode is expected to completely shield the electromagnetic waves emitted from the stacked IJJs.
Since the superconducting electrodes do not help thermal homogeneity, their results rather support the scenario (a).
Therefore, we conclude that the synchronized Josephson plasma oscillation is not excited in the mesas with $t_e=$ 400 nm because of the lack of inhomogeneity to induce the $\pi$-kinks aligned along the $c$-axis.

In summary, the electrode thickness on the top of the mesa is a key factor for THz emission from the Bi2212 intrinsic Josephson junction.
Intensive THz emission was observed from mesas with the electrode thickness less than 100 nm, in which remarkable temperature inhomogeneity is expected.
We claim that the local temperature rise inside the mesa induces the synchronization of the Josephson plasma oscillations of the large number of stacked intrinsic Josephson junctions.


\begin{acknowledgments}
The authors acknowledge T. Kashiwagi, M. Tsujimoto, T. Kawasaki, N. Hirayama, and T. Nakagawa for their technical assistance.
Critical comments from T. Koyama and X. Hu are also greatly appreciated.
This work was supported by the Mazda Foundation, the Sumitomo Foundation, and KAKENHI (23681030).
\end{acknowledgments}

\bibliography{THz_BIB,library}

\end{document}